\documentclass[12pt]{article}

\usepackage{graphicx}

\begin{document}

\title{Shape Phase Evolution of the Axially Symmetric States Between the
U(5) and SU(3) Symmetries}

\author{{Liang-zhu Mu$^{1}$, and Yu-xin Liu$^{1,2,3,4,}$\thanks{corresponding author} }\\[3mm]
\normalsize{$^1$ Department of Physics, Peking University, Beijing
100871, China}\\
\normalsize{$^2$ Key Laboratory of Heavy Ion Physics,
Ministry of Education, Beijing 100871, China}\\
\normalsize{$^3$ Institute of Theoretical Physics, Academia
Sinica, Beijing 100080, China}\\
\normalsize{$^4$ Center of Theoretical Nuclear Physics, National
Laboratory } \\
\normalsize{of Heavy Ion Accelerator, Lanzhou 730000, China} }

\date{\today}

\maketitle

\begin{abstract}
The shape phase structure and its transition of the nucleus in the
transitional region between the U(5) and SU(3) symmetries is
restudied in the framework of coherent-state theory with angular
momentum projection in IBM-1. The certain angular momentum (or
rotation-driven ) effect on the nuclear shape is discussed. A
coexistence of prolate and oblate shapes is found for the ground
states of the transitional nuclei. A phase diagram in terms of the
deformation parameter and angular momentum is given.
\end{abstract}




\newpage


The nuclear shape phase transition has been one of the most
interesting and significant subject in the research of nuclear
structure (see for example Refs.\cite{ Alhassid867,Iachello87,
Iachello98,Jolie99,CKZ99,Iachello01,SO01,Cejnar02,Cejnar031,
Regan03,Iachello03,LG03,Cejnar032,Pan03,Iach04,Cejnar04,SO04}).
Then the ground state shape phase transition between the U(5) and
SU(3) symmetries of the interacting boson model-1 (IBM-1) has been
well studied
\cite{Iachello87,Iachello98,Jolie99,Iachello01,Pan03}. The study
using coherent state theory shows that there is a first order
phase transition from spherical shape to axially deformed shape at
a certain critical value of the control parameter
\cite{Iachello87} and there may involve coexistence of spherical
and axially deformed shapes \cite{Iachello98}. However, the result
of using coherent state theory with angular momentum projection
shows that there is no shape coexistence and the transition is
continuous as the control parameter changes\cite{Jolie99}. Even
so, the minimum corresponding to the negative deformation
parameter $\beta$ has not yet been paid attention. On the other
hand, an analytical solution, namely the X(5) symmetry, has
recently been found for the states around the critical point of
the phase transition from the U(5) symmetry to SU(3) symmetry
using the collective model \cite{Iachello01}, where the potential
is in an infinite square well about the shape variable $\beta$
($\beta$-soft). In this paper, we will restudy the ground state
shape phase transition and the shape coexistence using the
coherent state formalism with angular momentum projection. On the
other hand, it has been found in experiment that rotation (or
angular momentum) may induce a shape phase transition from the
spheroid to the axial ellipsoid\cite{Regan03}. Since the angular
momentum projection can extend the coherent state formalism to
study the shape of excited states with certain angular momentum
and has been successful in describing the symmetry of the critical
point in the transition from U(5) symmetry to O(6)
symmetry\cite{LG03}, we will also study the rotation-driven shape
evolution along the ground state band of the transitional nuclei
in the region of U(5)-SU(3) symmetries in this paper.


To study the shape phase transition, one takes usually the
parametrized Hamiltonian in the framework of the IBM-1
\begin{equation}\label{eqn:hamiltonian1}
 \hat{H} = c \left ( \eta \hat{n}_d - \frac{1-\eta}{N}
         \hat{Q}(\chi) \cdot \hat{Q}(\chi) \right ),
\end{equation}
where $\hat{n}_d$ is the d-boson number operator and
$\hat{Q}(\chi) = [s^{\dagger} \tilde{d} + d^{\dagger} s ]^{(2)} +
\chi [d^{\dagger} \tilde{d}]^{(2)}$ is the quadrupole operator.
The $c$ just scales the interaction strength. The $N$ in the
denominator is the total boson number\cite{Jolie99,Cejnar98}.
Eq.(\ref{eqn:hamiltonian1}) can also be written as
\begin{eqnarray}\label{eqn:hamiltonian2}
 \hat{H} & = & \left[ \eta - \frac{2(1-\eta)\chi}{7N}\Big(\chi
 + \frac{\sqrt{7}}{2} \Big) \right] c C_{1U(5)} - \frac{2(1-\eta)
 \chi}{7 N} \Big( \chi + \frac{\sqrt{7}}{2} \Big) c C_{2U(5)}
 \nonumber \\
 & &   - \frac{2(1-\eta)}{\sqrt{7} N} \Big( \chi + \frac{\sqrt{7}}{2}
 \Big) c C_{2O(6)} + \frac{2(1-\eta)(\chi + \sqrt{7})}{7 N}
 \Big( \chi + \frac{\sqrt{7}}{2} \Big) c C_{2O(5)} \nonumber \\
 & &  + \frac{(1-\eta)\chi}{\sqrt{7}N} c C_{2SU(3)}
- \frac{(1-\eta)\chi (\chi + 2 \sqrt{7})}{14 N} c C_{2O(3)} \, ,
\end{eqnarray}
where $C_{kG}$ is the $k$-rank Casimir operator of group $G$. It
is evident that, at certain points of the parameter space, the
above Hamiltonian reaches the dynamical symmetry limits of the
IBM-1. If $\eta=1$ and $\chi$ arbitrary, it gives the U(5)
symmetry without two-body interactions being included; if $\eta=0
$ and $\chi=-\sqrt 7/2$, it is in the SU(3) symmetry; if $\eta=0 $
and $\chi=0$, it possesses the O(6) symmetry. Meanwhile, one can
realize the transition from the U(5) symmetry to the SU(3)
symmetry by fixing $\chi = -\frac{\sqrt{7}}{2}$ and varying $\eta$
from $1$ to $0$.

To study the geometric content of the system with Hamiltonian in
Eq.(\ref{eqn:hamiltonian1}), we take the intrinsic coherent state
$|g \rangle$ \cite{Iachello87}
\begin{equation} \label{eqn:coherentstate}
 |g \rangle  = (b^\dagger_g)^N |0 \rangle ,
\end{equation}
with  $$ b^{\dagger} _{g} = s^{\dagger} + \beta \cos \gamma
d^{\dagger}_0 + \frac{1}{\sqrt{2}}\beta \sin \gamma
(d^{\dagger}_{-2} + d^{\dagger}_{2})\, , $$
 where $s^{\dagger}$, $d^{\dagger}_{\mu}$ are the boson creation operators,
$|0 \rangle$ is the boson vacuum, and $\beta, \gamma$ are the
Hill-Wheeler intrinsic variables. It has been shown that the
$\beta$ and $\gamma$ here are proportional to the deformation
parameters $\hat{\beta}_{2}$ and $\hat{\gamma}$ in the collective
model (e.g., for rare earth nuclei, $\hat{\beta}_{2} \approx 0.15
\beta$ and $\hat{\gamma} = \gamma$)\cite{Iachello87}. Therefore
one usually refers the $\beta$ and $\gamma$ here to deformation
parameters for simplicity. Since the U(5) and SU(3) symmetries in
classical limit correspond to spherical and axially deformed
shapes\cite{Iachello87}, respectively, we can take only the
axially symmetric case, where $\gamma = 0$ or $\gamma = \pi$, into
account. To the convenience of description, we take $\gamma = 0$
and let $\beta$ be able to be positive or negative, corresponding
to the prolate or oblate shape, respectively\cite{Cejnar032}. Then
we can obtain the equilibrium shape of the ground state by
evaluating the potential energy surface $E(N,\beta )$ at the above
intrinsic coherent state and minimizing the $E(N,\beta )$ with
respect to $\beta$. After some derivation, $E(N,\beta )$ can be
given as
\begin{eqnarray} \label{eqn:unprojected}
E(N,\beta ) &=& \frac{c N}{(1+\beta^2)^2} \left\{
\frac{5(\eta-1)}{N}
      + \left[ (5\eta-4) + \frac{(\eta-1)(2+\chi^2)}{N} \right] \beta^2
      \right. \nonumber\\
  & & \left. + 4\sqrt{2} (N\!-\!1) \frac{(1 \! - \! \eta)\chi}{\sqrt{7}N}\beta^3
    + \frac{7[(N\!+\!1)\eta \!-\!1]\! + \!(5\! +\! 2N)(\eta\!-\!1) \chi^2 }{7N}
       \beta^4   \right\} \, .
\end{eqnarray}

In order to obtain the potential energy surface of the nuclear
states with certain angular momentum $L$, we take the angular
momentum projection on the coherent
state\cite{Lamme68,Dobes85,Kuyucak87}. Generally the angular
momentum projection operator is written
as\cite{Dobes85,Kuyucak87,HS95}
\begin{equation}
P^L_{MK} = \frac{2L+1}{8\pi^2}\int
D^{*L}_{MK}(\Omega)R(\Omega)d\Omega \, ,
\end{equation}
where $R(\Omega)$ is the rotational operator, $D^L_{MK}(\Omega)$
is the rotational matrix, and $\Omega$ is the Euler rotational
angle $(\alpha',\beta',\gamma')$. Taking advantage of the
phenomenological IBM, one knows that the states in the ground
state band are the ones with $Z$-component $K=0$ in the intrinsic
frame. Then to study the ground state band, we limit the
$P^{L}_{MK}$ to $P^{L}_{00}$. The potential energy functional of
the states in the ground state band (or the yrast band) can thus
be expressed as
\begin{equation}
E(N,L,\beta)=\frac{\left< g \right|HP^L_{00} \left| g
\right>}{\left< g \right|P^L_{00} \left| g \right>} \, .
\end{equation}
 With the explicit form of $R(\Omega)$ being substituted into, we
obtain the potential energy functional (in fact, since what we
consider at present is only the case at zero temperature, such a
potential energy functional is just the free energy) as
\begin{equation}\label{eqn:projected}
E(N,L,\beta) =  \frac{\int d \beta' \sin \beta' d^L_{00}(\beta')
\left< g \right|\hat H e^{-i\beta' J_y} \left| g \right>}{\int d
\beta' \sin \beta' d^L_{00}(\beta') \left< g \right|e^{-i\beta'
J_y} \left| g \right>} \, ,
\end{equation}
where $d^L_{00}(\beta')$ is the reduced rotational matrix. Noting
that $d^L_{00}(\beta')=P_L(\cos \beta')$, where $P_L$ is the
Legendre polynomial, we can accomplish the integrations and obtain
the explicit expression of the free energy numerically. Meanwhile,
It is easy to find that $E(N,L,\beta)=0$ if $L>2N$. Such a result
is consistent with what the phenomenological IBM gives. Therefore
we take the angular momentum $L \in [0,2N]$ in the following
discussion. To discuss only the axially symmetric states in the
transitional region between the U(5) and SU(3) symmetries, we vary
the parameter $\eta$ and fix the angular deformation parameter
$\gamma \equiv 0$ and the parameter $\chi = - \frac{\sqrt{7}}{2}$.

\begin{figure}
\begin{center}
\includegraphics[scale=0.8]{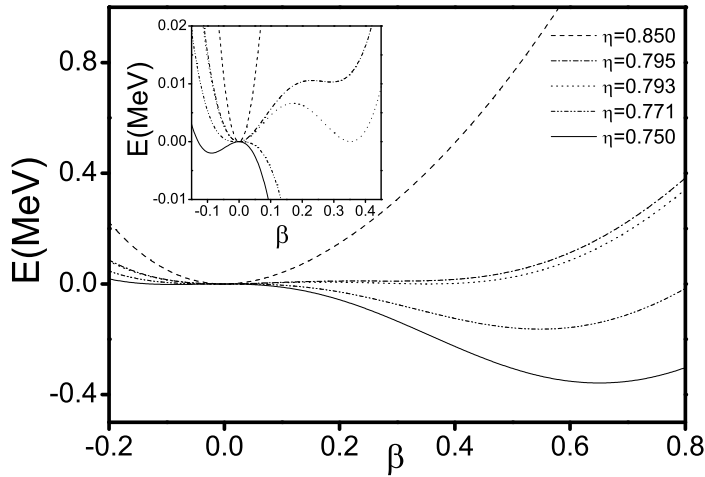}
\caption{\label{fig:unprojected} Potential energy surface
$E(N,\beta)$ for the nuclei with total boson number $N=10$. Five
curves are plotted for different values of the control parameter
$\eta$ as the same as those in Ref.\cite{Jolie99}.}
\end{center}
\end{figure}
\begin{figure}
\begin{center}
\includegraphics[scale=0.8]{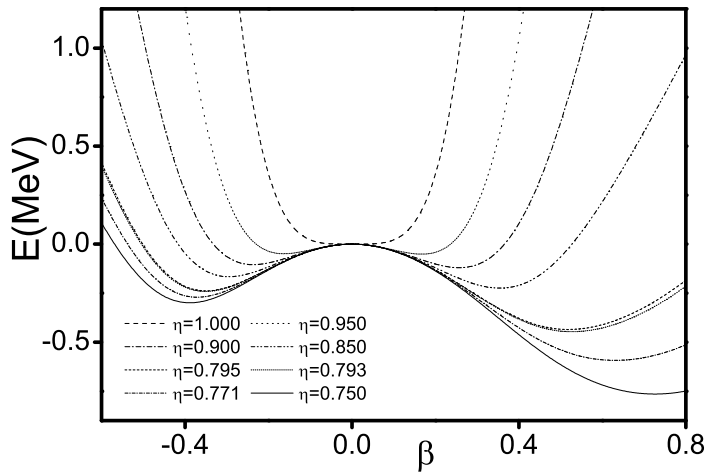}
\caption{\label{fig:projected} Effect of varying the control
parameter $\eta$ on the potential energy surface $E(N,L,\beta)$
for the nuclei with total boson number $N=10$ and angular momentum
$L=0$.}
\end{center}
\end{figure}

Ref.\cite{Jolie99} has taken the nucleus with total boson number
$N=10$ as an example to study the ground state shape phase
transition using Eq.(\ref{eqn:unprojected}). It shows that the
potential energy surface involves a double minimum structure,
corresponding to spherical and axially deformed shapes,
respectively, in a very narrow region of control parameter, namely
$0.771\leq \eta \leq0.795$\cite{Iachello98, Jolie99}. Then a first
order shape phase transition happens when the global minimum
changes discontinuously from $\beta=0$ (spherical shape) to
$\beta>0$ (deformed shape) at $\eta = 0.793$. Because the barrier
between the two minima is shallow, the coexistence of spherical
and deformed shapes can appear. Our restudy at present reproduces
this result well, which is illustrated in the $\beta \geq 0$ part
of Fig.~\ref{fig:unprojected}. As the certain angular momentum
effect is involved by taking the angular momentum projection, the
study in Ref.\cite{Jolie99} using Eq.(\ref{eqn:projected}) shows
that no double minimum structure is present and the global minimum
shifts continuously from $\beta=0$ for $\eta=1$ to $\beta>0$ for
$\eta<1$, whereas a local maximum develops at $\beta=0$. Our
restudy reproduces such a result well, too, as shown in the $\beta
\geq 0$ part of Fig.\ref{fig:projected}.

However, the above study is limited to $\beta \geq 0$. In fact,
the value of the deformation parameter $\beta$ can be negative,
which corresponds to a oblate shape. So it is necessary to include
the case of $\beta<0$ to give a complete understanding of shape
phase structure and its transition of the nuclei in the between of
U(5) and SU(3) symmetries. Then we show the complete feature of
the potential energy surface expressed in
Eqs.(\ref{eqn:unprojected}) and (\ref{eqn:projected}) in
Figs.\ref{fig:unprojected} and \ref{fig:projected}, respectively.
From Fig.\ref{fig:unprojected}, one can recognize that the results
does not change except that, after the local minimum at $\beta =
0$ becomes a local maximum, a local minimum at $\beta < 0$
develops and changes continuously as $\eta$ decreases. Since the
minimum at $\beta > 0$ is much lower than that at $\beta <0$, the
coexistence of prolate and oblate shapes is not likely to appear.
The result illustrated in Fig.\ref{fig:projected}, however, shows
that, if the control parameter $\eta > 0.850$, the difference of
the two minima corresponding to $\beta >0$, $\beta <0$,
respectively, is quite small and the barrier between the minima is
shallow. Then the prolate and oblate deformation shapes may
coexist. Actually, as the control parameter $\eta \rightarrow 1$,
the values of $E(N,L=0,\beta)$ at the two minima become more and
more close and coincide at $\eta = 1$, where $\beta = 0$. It seems
that no first order shape phase transition occurs, because the
global minimum at $\beta >0 $ changes continuously with the
control parameter $\eta$.

\begin{figure}
\begin{center}
\includegraphics[scale=0.8]{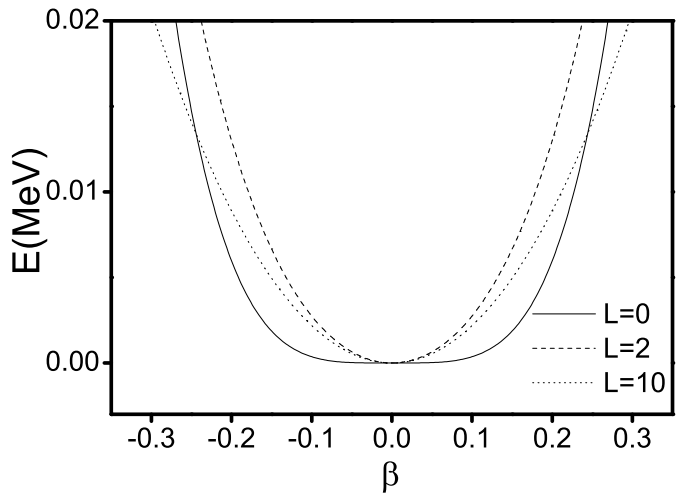}%
\caption{\label{fig:U5} The potential energy surface
$E(N,L,\beta)$ for the states in the U(5) symmetry with boson
number $N=10$ and angular momentums $L=0, 2, 10$.}
\end{center}
\end{figure}
\begin{figure}
\begin{center}
\includegraphics[scale=0.8]{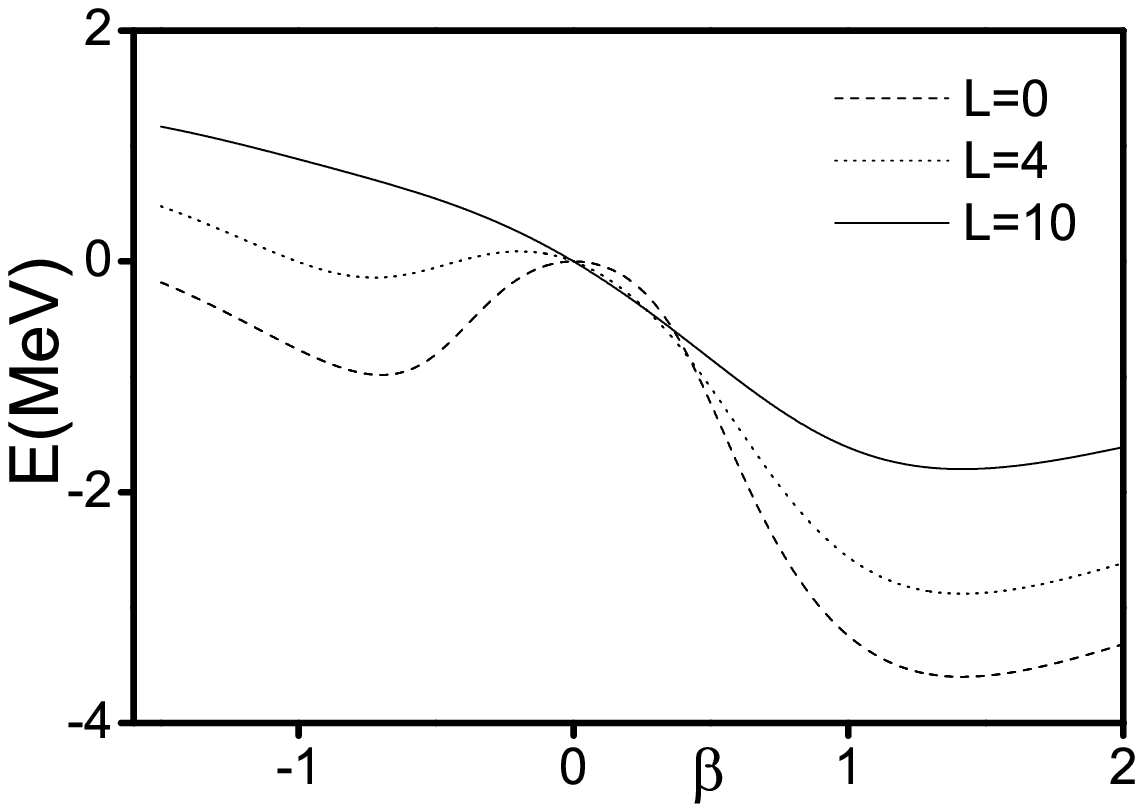}%
\caption{\label{fig:SU3} The potential energy surface
$E(N,L,\beta)$ of the states in the SU(3) symmetry with boson
number $N=10$ and angular momentums $L=0, 4, 10$.}
\end{center}
\end{figure}
\begin{figure}
\begin{center}
\includegraphics[scale=0.8]{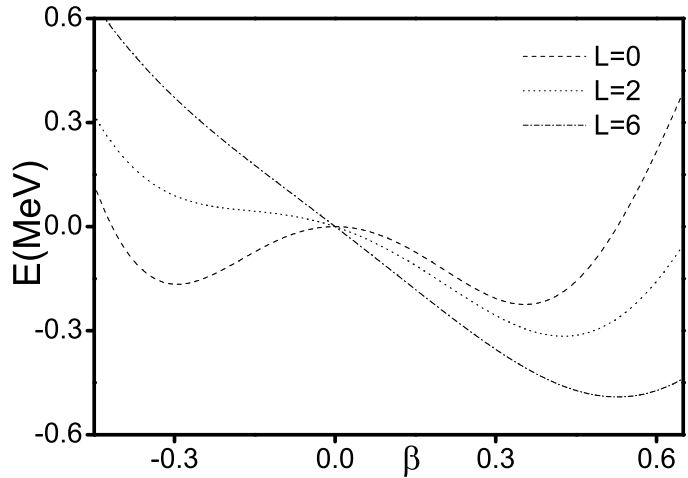}%
\caption{\label{fig:transitional}The potential energy surface
$E(N,L,\beta)$ for the boson number $N=10$ and angular momentum
$L=0, 2, 6$ states of the transitional nucleus with control
parameter $\eta = 0.85$. }
\end{center}
\end{figure}

With various angular momentum $L = 0, 2, 4, 6, \cdots$ being taken
in Eq.(\ref{eqn:projected}), we can study the rotation (angular
momentum) driven effect on the nuclear potential energy surface.
The results for the U(5) symmetry limit with the two-body
interactions being neglected (with $\eta =1$), the SU(3) symmetry
limit (with $\eta =0$) and the interplay between the U(5) and
SU(3) symmetries (taking the case with $\eta = 0.850$ as an
example) are illustrated in Figs.\ref{fig:U5}, \ref{fig:SU3},
\ref{fig:transitional}, respectively. Fig.\ref{fig:U5} shows
obviously that all the $E(N,L,\beta)$s with different angular
momentum $L$ have the same minimum at $\beta=0$. It means that
rotation does not change the structure of the potential energy
surface $E(N,L,\beta)$ of the states in U(5) symmetry without
two-body interactions, so that the states maintain vibrational
ones and the nucleus appears as a spheroid. However, the
$\beta$-soft\cite{Cejnar032} disappears as $E(N,L,\beta)$ becomes
steeper around $\beta =0$ if the angular momentum takes a nonzero
value. From Fig.\ref{fig:SU3}, one can recognize that, the
potential energy surface $E(N,L,\beta)$ of the states with low
angular momentum and SU(3) symmetry involves a global minimum at
$\beta = \sqrt{2}$ and a local minimum at $\beta < 0$.
Corresponding to the increase of angular momentum, the local
minimum at $\beta < 0$ gets shallowed, and then disappears. It
indicates that the stable shape of all the states in the SU(3)
symmetry always appears as an prolate ellipsoid. Furthermore, such
a prolate shape gets more stable and the metastable oblate shape
disappears with the angular momentum increasing. For the
transitional nuclei, for example the one with control parameter
$\eta = 0.850$,  from Fig.\ref{fig:transitional}, one can infer
that the prolate and oblate shapes may coexist at the state with
angular momentum $L=0$, since the values of the two minima at
$\beta > 0$ and $\beta < 0$ are quite close to each other. As the
angular momentum increases, the minimum at $\beta < 0$ becomes
shallower and eventually disappears, meanwhile the minimum at
$\beta > 0$ gets deeper and the corresponding $\beta$ increases.
It indicates that rotation breaks the shape coexistence and makes
the deformation of the stable shape bigger and bigger. As a
critical angular momentum is reached, the nucleus appears only in
a prolate ellipsoid shape. If the shape coexistence can be taken
as a special shape phase, we can say that a shape phase transition
from coexistence to prolate shape occurs. And the phase transition
may be continuous.

\begin{figure}[ht]
\begin{center}
\includegraphics[scale=0.8]{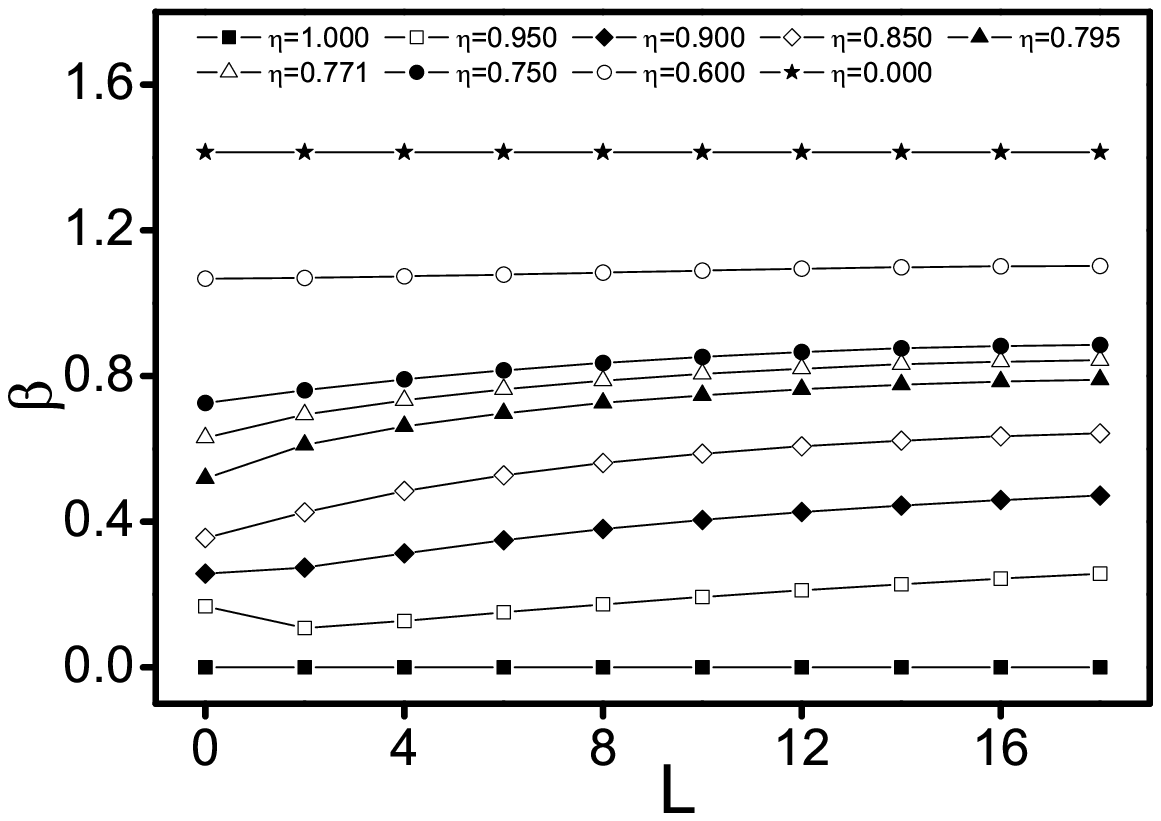} %
\caption{\label{fig:angl-beta} The phase diagram in terms of the
deformation parameter $\beta$ and the angular momentum $L$ of the
transitional nuclei in the between of U(5) and SU(3) symmetries. }
\end{center}
\end{figure}

To show the characteristic of the shape evolution of the
transitional nuclei in the region from U(5) symmetry to SU(3)
symmetry against the angular momentum $L$ more clearly, we
illustrate the phase diagram in terms of the deformation parameter
and angular momentum at several values of the control parameter
$\eta$ in Fig.\ref{fig:angl-beta}. From Fig.\ref{fig:angl-beta}
and Fig.\ref{fig:SU3}, one may conclude that the nucleus with
SU(3) symmetry is like a rigid rotor, the shape of which is so
hard as not to be affected by the rotation and without any
vibrational freedom. Then the energy spectrum of SU(3) symmetric
nucleus is in good rotational. By the way, it should be mentioned
that the deformation parameter $\beta$ ( $= \sqrt{2}$) of the
states in SU(3) symmetry and with certain angular momentum is just
the same as that of the classical limit (total boson number $ N
\rightarrow \infty$) in the case of without certain angular
momentum. Whereas, the nucleus with U(5) symmetry is like a
spheroid, and such a shape is not influenced by the rotation,
either. Therefore the spectrum of the nucleus in U(5) symmetry is
in good vibrational. For the transitional nucleus in the between
of U(5) and SU(3) symmetries, it is like a soft liquid drop, whose
shape may change to prolate ellipsoid and the deformation keeps
increasing as the rotation gets rapid. Such an increase of
deformation parameter with the increasing angular momentum is a
corroboration of the well known variable moment of
inertia\cite{MSB69}. Looking over Fig.\ref{fig:angl-beta} more
carefully, one may infer that, as the control parameter $\eta$
takes a value about $0.8$, the deformation parameter $\beta$
changes with respect to the angular momentum most drastically.
Then it may be around the critical point between the U(5) symmetry
and the SU(3) symmetry. Referring Eq.(\ref{eqn:hamiltonian1}), one
knows that, the ratio between the interaction strength with U(5)
symmetry (without two-body interactions being taken into account)
and that with SU(3) symmetry $\frac{\eta}{(1-\eta)/N} \approx
\frac{0.8}{0.2/10}=40$. Recalling the results given in
Refs.\cite{Pan03} and \cite{Werner02}, such a ratio is just that
of the critical symmetry X(5). Combining this point with
Fig.\ref{fig:transitional}, one can infer that the ground state of
the nucleus around the critical point may involve prolate and
oblate shape coexistence, and its shape is so soft that it changes
easily with the increase of angular momentum. It is also
remarkable that, for the states differing from those with U(5)
symmetry very slightly (e.g., $\eta = 0.95$), the one with angular
momentum $L=2$ involves a smaller deformation parameter than its
neighbors (e.g., with $L = 0, 4$). Since a smaller deformation
parameter means a smaller rotational moment of inertia and
contributes a relatively larger anharmonic effect (in rotation),
such a phenomenon may be a manifestation of the $2^{+}_{1}$
anomaly of near-vibrational states\cite{KJ9571}. Up to now, a lot
of works have been developed to remove the anomaly and to explore
the mechanism(see for example Ref.\cite{CC02} and references
therein), the underlying physics remains to be studied further.


In conclusion, we have studied the shape phase structure and its
evolution of the axially symmetric nuclear states in the between
of the U(5) and the SU(3) symmetries in IBM-1. Especially the
certain angular momentum effect or the rotation driven effect is
studied by taking the angular momentum projection on the intrinsic
coherent state. A phase diagram in terms of the deformation
parameter and angular momentum is given. Meanwhile the case of
$\beta < 0$, which corresponds to a oblate deformed shape, is
included. The results show that there still can be a shape
coexistence as the certain angular momentum effect is taken into
account. However the coexistence is of oblate and prolate deformed
shapes, but not of spherical and prolate shapes. Meanwhile, as the
certain angular momentum effect is considered, the nucleus with
SU(3) symmetry always appears as a rigid rotor with deformation
parameter taking the classical limit value $\sqrt{2}$, and the one
with U(5) symmetry as a pure vibrator. The nucleus in the
transitional region between the U(5) symmetry and the SU(3)
symmetry appears as a soft liquid drop, which involves both
rotational and vibrational degrees of freedom. Then the ground
state of some transitional nuclei may involve coexistence of
prolate and oblate shapes. With the increasing of the angular
momentum, the stable shape can only be prolate ellipsoid, whose
deformation can get larger and larger.  It also suggests that the
states with the critical symmetry X(5) may be the one with prolate
and oblate shape coexistence.  Finally, it is remarkable that, the
Hamiltonian we used here is very simple, the real case may be much
more complicated. The related investigation is now under progress.

\bigskip


This work was supported by the National Natural Science Foundation
of China under contract Nos. 10425521, 10075002, 10135030, the
Major State Basic Research Development Program under contract No.
G2000077400 and the Doctoral Program Foundation of the Ministry of
Education, China, under grant No 20040001010. One of the authors
(Y.X. Liu) would also acknowledge the support of the Foundation
for University Key Teacher by the Ministry of Education, China.


\newpage


\begin{thebibliography}{99}
\bibitem{Alhassid867} Y. Alhassid, S. Levit, and J. Zingman, Phys. Rev.
Lett. {\bf 57} (1986), 5639; Y. Ahlassid, J. Zingman, and S.
Levit, Nucl. Phys. {\bf A 469} (1987), 205.
\bibitem{Iachello87} F. Iachello and A. Arima, \emph{The Interacting Boson Model}
(Cambridge University Press, Cambridge, 1987).
\bibitem{Iachello98} F. Iachello, N. V. Zamfir, and R. F. Casten,
Phys. Rev. Lett. \textbf{81} (1998),1191.
\bibitem{Jolie99} J. Jolie, P. Cejnar, and J. Dobe\v{s}, Phys. Rev.
\textbf{C 60} (1999), 061303(R).
\bibitem{CKZ99} R. F. Casten, D. Kusnezov, and N. V. Zamfir, Phys. Rev.
Lett. {\bf 82} (1999), 5000.
\bibitem{Iachello01} F. Iachello, Phys. Rev. Lett. \textbf{87} (2001), 052502.
\bibitem{SO01} N. Shimizu, T. Otsuka, T. Mizusaki, M. Honma, Phys. Rev.
Lett. {\bf 86} (2001), 1171.
\bibitem{Cejnar02} P. Cejnar, Phys. Rev. {\bf{C 65}} (2002), 044312.
\bibitem{Cejnar031} P. Cejnar P, Phys. Rev. Lett. {\bf{90}} (2003), 112501.
\bibitem{Regan03} P. H. Regan, {\it{et al.}}, Phys. Rev. Lett. {\bf
90} (2003), 152502.
\bibitem{Iachello03} F. Iachello, Phys. Rev. Lett. {\bf{91}} (2003), 132502.
\bibitem{LG03} A. Leviatan and J. N. Ginocchio, Phys. Rev. Lett. {\bf{90}}
(2003), 212501.
\bibitem{Cejnar032}  P. Cejnar, S. Heinze, and J. Jolie, Phys. Rev. \textbf{C 68}
(2003), 034326.
\bibitem{Pan03} F. Pan, J. P. Draayer, and Y. A. Luo, Phys. Lett.
\textbf {B 576} (2003), 297.
\bibitem{Iach04} F. Iachello, and N. V. Zamfir, Phys. Rev. Lett.
{\bf{92}} (2004), 212501.
\bibitem{Cejnar04} P. Cejnar and J. Jolie, Phys. Rev. {\bf{C 69}} (2004),
011301(R).
\bibitem{SO04} N. Shimizu, T. Otsuka, T. Mizusaki, M. Honma, Phys. Rev.
{\bf C 70} (2004), 054313.
\bibitem{Cejnar98}  P. Cejnar and J. Jolie, Phys. Rev. \textbf{E 58}
(1998), 387.
\bibitem{Lamme68}   H. A. Lamme and E. Boeker, Nucl. Phys. {\bf A 111}
(1968), 492.
\bibitem{Dobes85}   J. Dobe\v{s}, Phys. Lett. {\bf B 158} (1985), 96;
Phys. Rev. {\bf C 42} (1990), 2023.
\bibitem{Kuyucak87} S. Kuyucak and I. Morrison, Phys. Rev. {\bf C 36}
(1987), 774.
\bibitem{HS95} K. Hara, and Y. Sun, Int. J. Mod. Phys. {\bf E 4} (1995), 637.
\bibitem{MSB69} M.A.J. Mariscotti, G. Scharff-Goldhaber, and B.
Buck, Phys. Rev. {\bf 178} (1969), 1864.
\bibitem{Werner02} V. Werner, P. von Brentano, R.F. Casten, and J. Jolie,
Phys. Lett. {\bf B 527} (2002), 55.
\bibitem{KJ9571} J. Kern, P. E. Garrett, J. Jolie, and H. Lemann,
Nucl. Phys. {\bf A 593} (1995), 21; J. Kern, and J. Jolie, Phys.
Lett. {\bf B 364} (1995), 207; J. Kern, and J. Jolie, Nucl. Phys.
{\bf A 624} (1997), 415; P. Cejnar, J. Jolie, and J. Kern, Phys.
Rev. {\bf C 63} (2001), 047304.
\bibitem{CC02} M. A. Caprio, and R. F. Casten, and J. Jolie, Phys.
Rev. {\bf C 65} (2002), 034304.

\end{thebibliography}

\end{document}